\documentclass[prl,preprint,showpacs,amsmath]{revtex4-1}

\usepackage{amssymb}
\usepackage[caption=false]{subfig}
\usepackage{psfrag}

\newcommand{\Ab}{\textbf{A} }

\newcommand{\pb}{\textbf{p} }
\newcommand{\rb}{\textbf{r} }
\newcommand{\vb}{\textbf{v} }

\newcommand{\sgn}{\ensuremath{\operatorname{sgn}}}

\newcommand{\g}{\ensuremath{\gamma} }
\newcommand{\eps}{\ensuremath{\epsilon} }

\newcommand{\ome}{\ensuremath{\omega_{pe}} }

\newcommand{\ct}{\ensuremath{\tilde{c}} }

\newcommand{\tpxx}{\ensuremath{\tilde{p}_{||}}}
\newcommand{\tpyy}{\ensuremath{\tilde{p}_\bot}}
\newcommand{\tyy}{\ensuremath{\tilde{\rho}}}

\begin{document}

\title{Temporal and spatial expansion of a multidimensional model for electron acceleration in the bubble regime}
\author{Johannes Thomas}\email{thomas@tp1.uni-duesseldorf.de}
\author{Alexander Pukhov}
\affiliation{Institut f\"ur Theoretische Physik I, Heinrich-Heine-Universit\"at D\"usseldorf, D-40225 Germany}

\begin{abstract}
An extended analytical model for particle dynamics in fields of a highly-nonlinear plasma wake field (the bubble or blow out regime) is derived. A recently proposed piecewise model (Kostyukov et al., New J. Phys., {\bf 12 }, 045009 (2010)) is generalized to include a time dependent bubble radius and full field solution in the acceleration direction. Incorporation of the cavity dynamics in the model is required to simulate the particle trapping properly. On the other hand, it is shown that the previously reported piecewise model does not reproduce the formation of a mono energetic peak in the particle spectrum. The mono energetic electron beams are recovered only when the full longitudinal field gradient is included in the model. 
\end{abstract}

\maketitle

When a plasma wake field is driven to nonlinear amplitudes, it takes the form of an electronic cavity that is called "bubble" in the case of a short pulse laser driver \cite{Pukhov2002} or "blowout regime" in the case of dense highly energetic particle bunches \cite{Rosenzweig1991, Lotov2004}. For the laser driven case, the pulse  is shorter than the plasma wavelength and fits perfectly into the first half of the plasma period. The laser intensity is thought to be high enough that the created wake field breaks after its first oscillation. In this regime, the wake field takes the form of a distorted spherical cavity from which all electrons are banished and that moves with nearly the speed of light through the plasma. In the following we refer this wake-field  as "the bubble".

Inside the bubble, an accompanying electron bunch - the so called beam load - is accelerated until the spherical shape of the bubble breaks down or the electrons get out of phase with the wake. The break down of the sphere happens as soon as the driver energy is exhausted \cite{Tajima1979, Chen1985}. The fields inside the bubble can be calculated from Maxwell's equations for an empty sphere \cite{Kostyukov2004} and a sphere that takes into account the field enhancement at the bubble back due to electron sheet crossing \cite{Bulanov1997,Lu2007}. An even  simpler model for the fields inside the bubble has been proposed recently:  a static piecewise model (PWM) \cite{Kostyukov2010}. This simplified model neglects any spatial or temporal field dependence, but due to its simplicity it is possible to calculate amplitude envelope functions to the transverse particle coordinates and momenta, and to analyze some self-injection physics. In contrast to recent experiments, the PWM misses the production of mono energetic electron beams. Thus, we are forced to modify the PWM to a more general case that remains analytically treatable and improves the self-injection physics found in the PWM.

Our first approach is to consider fields that are time dependent, but piecewise constant in space. In this case our model is a sphere with an adiabatically growing radius $R(t)$ and a growing trapping cross-section. A physical justification for this approach is that now the interaction of the electron bunch with the bubble border is taken into account. Since the fields remain constant we still miss the production of mono energetic electron beams. Thus, in a second step we introduce a model that treats the right field gradient in the longitudinal direction. For this case indeed we find energy spectra that are comparable to PIC simulations and a full gradient model.

Our work is based on the PWM. Thus, similar to \cite{Kostyukov2010}, our bubble models neglect the interaction between the accelerated electrons and the laser pulse at the bubble front. We also assume a continuous plasma background density, and a reduced electron dynamics in the z-$\rho$-plane, where $\rho$ is the perpendicular distance of an electron to the z-axis and z the propagation direction of the laser pulse. The motivation for all these assumptions are several observations in the  phenomenological model of the bubble regime \cite{Kostyukov2004, Kostyukov2009}.

The full gradient potentials inside a bubble are given in \cite{Kostyukov2004, Lu2006}. In normalized variables they are $A_{||} = -\varphi = \Phi/2$, and $\Ab_\bot = 0$, where $\Phi(\rb)=|\rb|^2/4$ is the bubble potential. The normalization is done by $\rb\rightarrow \rb\ome/c$ for length, $t\rightarrow t\ome$ for time, $\vb\rightarrow\vb/c$ for velocity, $\pb\rightarrow\pb/(m_ec)$ for the kinetic momentum, and $\Phi\rightarrow e\Phi/(m_ec^2)$ for potentials. Here $\ome=\left(4\pi e^2n_0/m_e\right)^{1/2}$ is the electron plasma frequency, $m_e$ is the electron mass, $c$ is the velocity of light, and $n_0$ is the plasma background density. The equations of motion of the full gradient model in a co-moving frame of reference are Eq.(1)-(2) in Ref.\cite{Kostyukov2009}. In the PWM, $\xi=z-Vt$ and $\rho$ are replaced by  
\begin{equation}
\xi = \sgn(\xi)F, \hspace{1cm} \rho = \sgn(\rho)F,	\label{eqn:F}
\end{equation} 
where $F$ is the field-strength parameter of the PWM and $V$ the group velocity of the laser pulse.\\

\textbf{The time-dependent PWM (tPWM)}\\
One possibility to examine the interaction between the electron bunch and the bubble border is to consider an adiabatically slow bubble growth. This growth significantly affects the electron self-injection physics because the bubble back moves slower than the bubble \cite{Kostyukov2004,Kostyukov2009,Whittum1992,Lu2007,Kalmykov2009,Yi2011,Kalmykov2011}.

To adopt the idea of a growing bubble to the tPWM we make the same consideration as in \cite{Kalmykov2009,Yi2011}. Applied to the full gradient model \cite{Kostyukov2004}, we find the sufficient trapping condition $\lim_{t\rightarrow\infty}\Phi(t)=-\infty$. Thus, a new approach for our wake field potential in the tPWM is
\begin{equation}
\Phi(\rb,t) = \frac{F}{2}(|\xi^-|+|\rho|) - \frac{R(t)^2}{4},
\end{equation}
with  $R(t) = (1+\eps t)R_0$, $ \xi^- = z-V^-t$, $V^-=V-\eps R_0$, $V=\sqrt{1-\g_0^{-2}}$, and the gamma-factor of the bubble front $\g_0$. In these coordinates the bubble center rests and the equations of motion for the tPWM are
\begin{eqnarray}
\frac{dp_{||}}{dt} & = & -\frac{1}{4}\Omega^- + \frac{\sgn(\rho)Fp_\bot}{4\g},													\label{eqn:tPWMEq_I}\\
\frac{dp_\bot}{dt} & = & -\left(1+\frac{p_{||}}{\g}\right)\frac{\sgn(\rho)F}{4},												\label{eqn:tPWMEq_II}\\
\frac{d\xi^-}{dt} 	 & = & \frac{p_{||}}{\g}-V^-, \hspace{0.5cm}	\frac{d\rho}{dt} = \frac{p_\bot}{\g},	\label{eqn:tPWMEq_III}
\end{eqnarray}
with $\Omega^- = (1+V^-)\sgn(\xi^-)F+(1+\eps t)R_0^2\eps$.

\begin{figure}[t]
	\centering
		\subfloat[]{\label{fig:paper_TIME_COMP_001}\resizebox{0.3\textwidth}{!}{\includegraphics{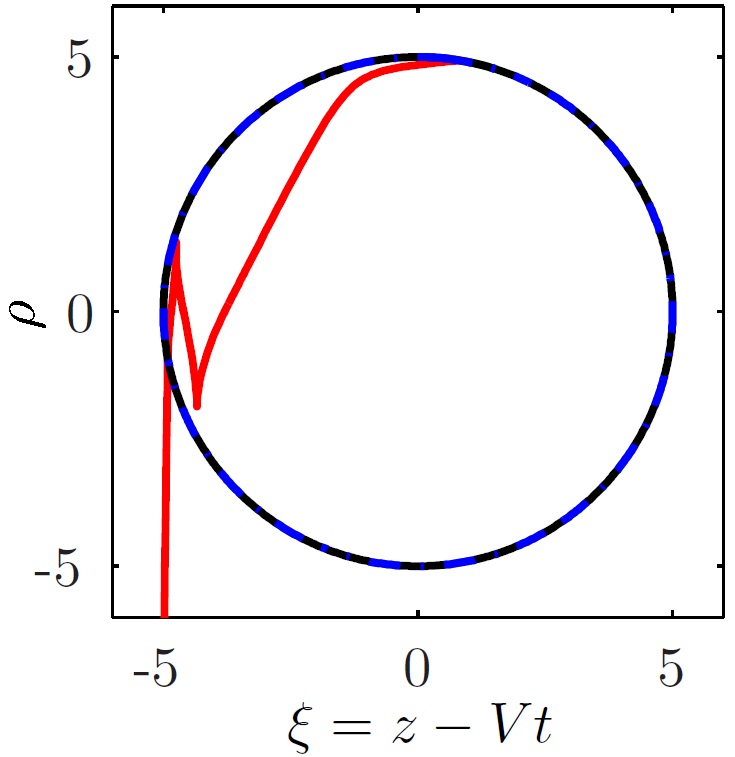}}}
		\hfill
		\subfloat[]{\label{fig:paper_TIME_COMP_M2}\resizebox{0.3\textwidth}{!}{\includegraphics{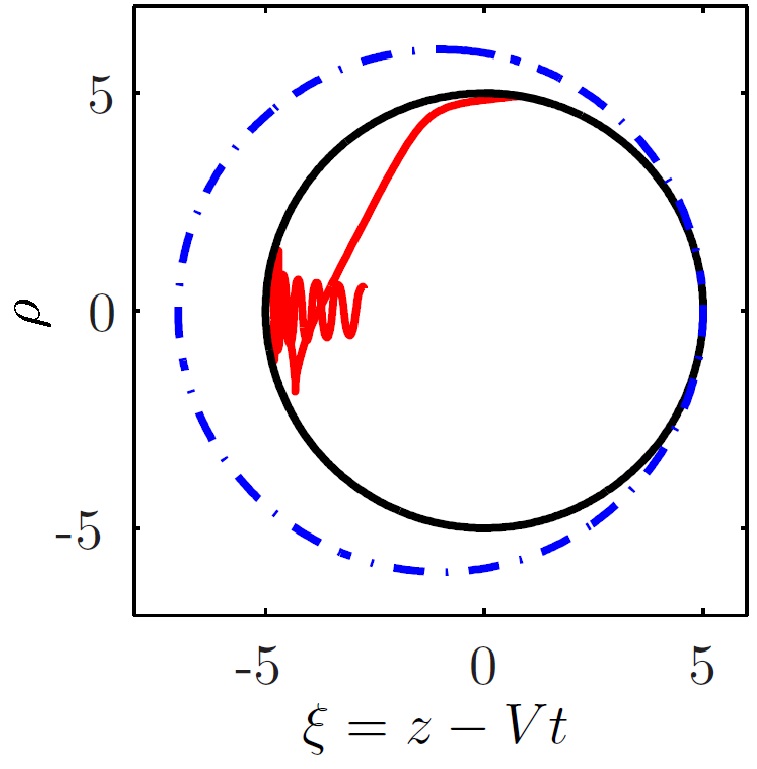}}}
		\hfill
		\subfloat[]{\label{fig:Grid_2D}\resizebox{0.3\textwidth}{!}{\includegraphics{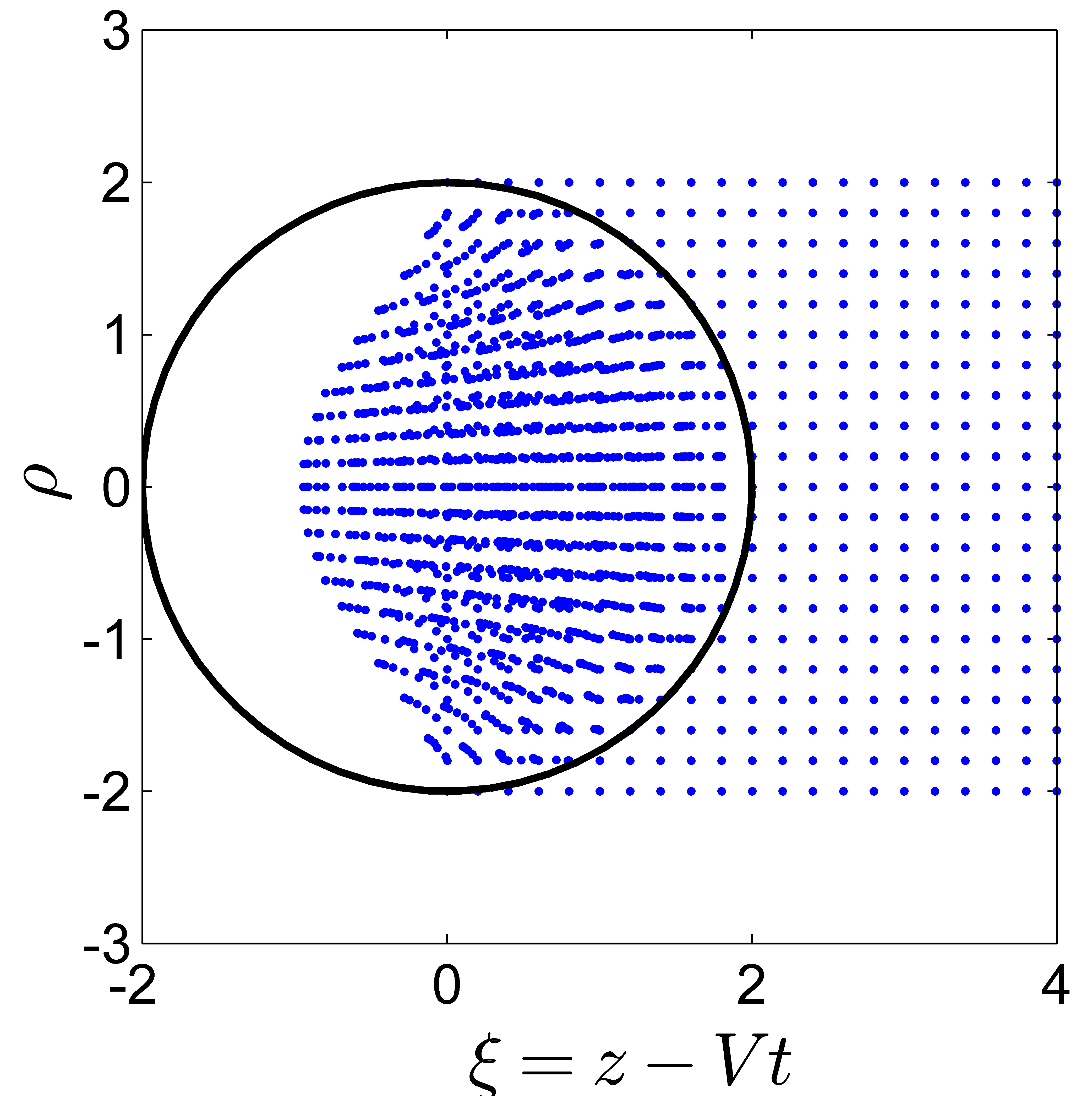}}}
	\caption{(\ref{fig:paper_TIME_COMP_001}): Trajectory in the PWM, particles is not trapped. (\ref{fig:paper_TIME_COMP_M2}): Trajectory in the tPWM, particle is trapped. Both simulations are done with $R_0=5$, $F=4$, $\rho_0 = 4.92$, and $\gamma_0=10$. The black solid circle has radius $R_0$. The blue dotted circle is the bubble after the simulation ends. (\ref{fig:Grid_2D}): Simulation configuration: Bubble running into a grid of electrons.}
	\label{fig:trapdyn}
\end{figure}

Based on the tPWM we perform two kinds of simulation. In the first part we solve the equations of motion (\ref{eqn:tPWMEq_I}-\ref{eqn:tPWMEq_III}) for a single electron and discuss its trajectory, while we always assume the initial conditions $(\xi_0^-)^2+\rho_0^2 = R_0^2$, $ \pb_0=0$, and $ \eps, \xi_0^-, \rho_0 >0$. In the second part we place a grid of particles with zero momentum in front of the bubble as shown in Fig.\ref{fig:Grid_2D} and solve the equations of motion for every individual particle. The resulting energy spectra of trapped particles then give us a better insight in the self injection dynamics and acceleration mechanism in the tPWM.

If we compare a particle's trajectory in both the PWM (Fig.\ref{fig:paper_TIME_COMP_001}) and the tPWM (Fig.\ref{fig:paper_TIME_COMP_M2}), we discover a new kind of trapping dynamic.
The particle in the static model is lost because its trajectory crossed the bubble border. In the time-dependent model, this situation can be prevented as the border moves out. The amplitude function of the oscillation in the trajectory determines whether the electron will be trapped. So,  Kostyukov et al. \cite{Kostyukov2010} describe a way to calculate the envelope approximation functions $\tilde{\rho}(t)$ and $\tilde{p}_\bot(t)$ to the amplitude functions of $\rho(t)$ and $p_\bot(t)$. We follow their steps in the limit $\eps R_0^2\ll 2F$ and introduce the times $t_n$ by $\rho(t_n)=0$ for all $n\geq1$. Then, with $n\gg1$ and a guiding center approximation to $p_{||}(t)$, we obtain  the equation system
\begin{eqnarray}
\tpxx(t)	& = & \frac{1}{4}[(1+V^-)F-R_0^2\eps] t,																									  	\label{eqn:tpxt}\\
\tpyy(t)	& = & 1.00\ct y_0^{2/3}\frac{F^{5/3}}{|\Omega^-(t)|}[(1+V^-)F-R_0^2\eps]^{1/3} t^{1/3},				\label{eqn:tpyt}\\
\tyy(t) 	& = & 4.01\frac{\ct^2\rho_0^{4/3}F^{7/3}}{\Omega^-(t)^2[(1+V^-)F-R_0^2\eps]^{1/3}} t^{-1/3},	\label{eqn:tyt}
\end{eqnarray}
while we use the constant $\ct=\ct(F,R_0,\g_0,\eps,\xi_0^-)$ to fit the curves to our simulations.

\begin{figure}[tbp]
	\centering
	\subfloat[]{\label{fig:paper_M2_460.20.000410_y_plot}\resizebox{5cm}{!}{\includegraphics{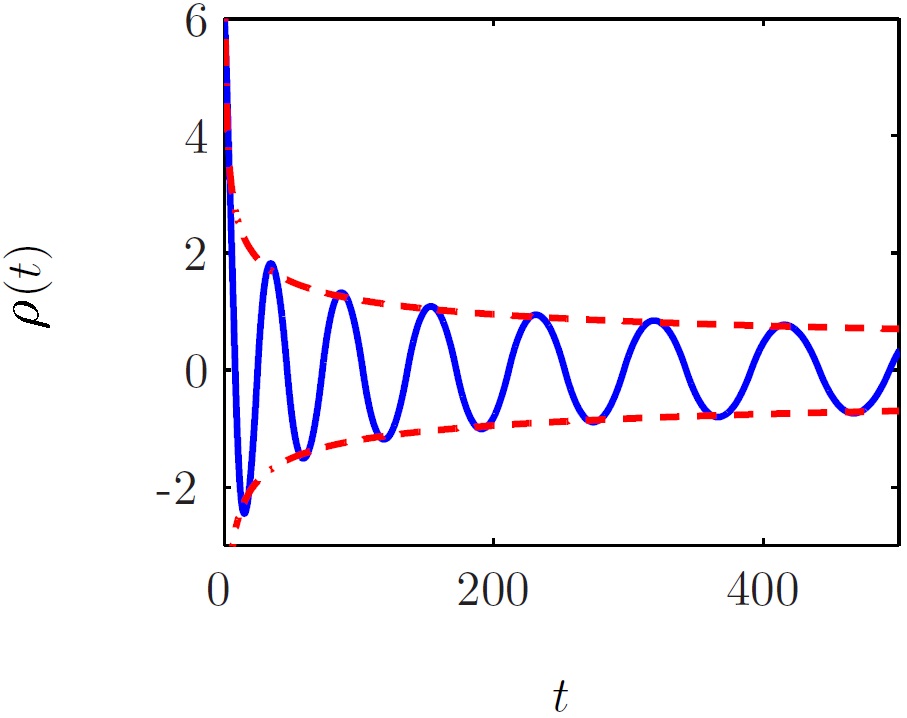}}}
	\hfill
	\subfloat[]{\label{fig:paper_M2_460.20.000410_py_plot}\resizebox{5cm}{!}{\includegraphics{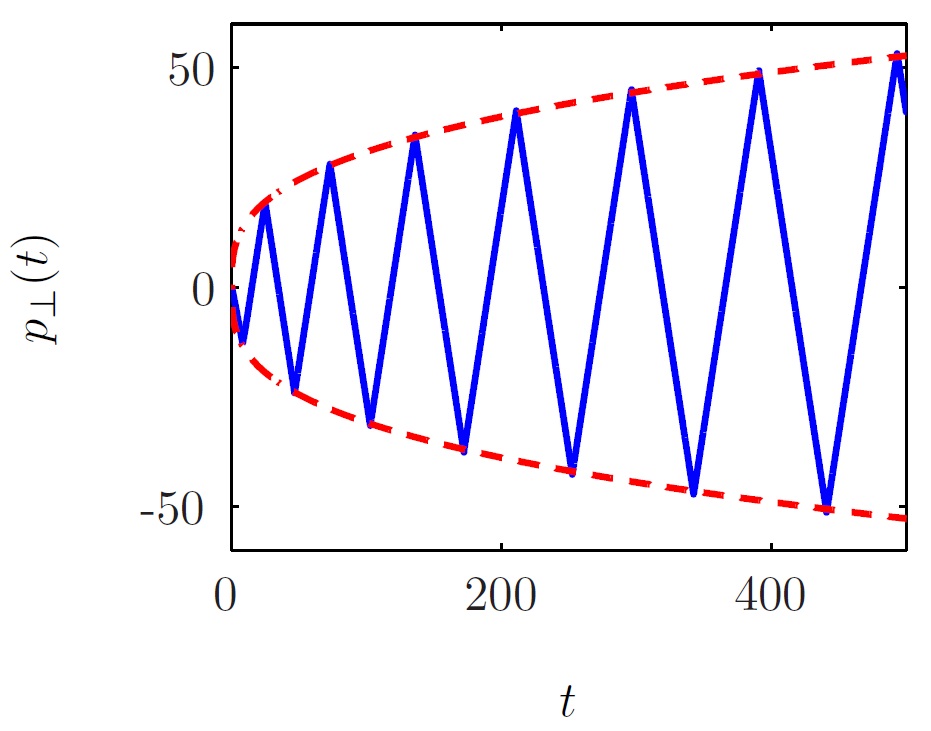}}}
	\hfill
	\subfloat[]{\label{fig:paper_MTC_M2_gamma=10_F=6_eps=0.0002}\resizebox{5cm}{!}{\includegraphics{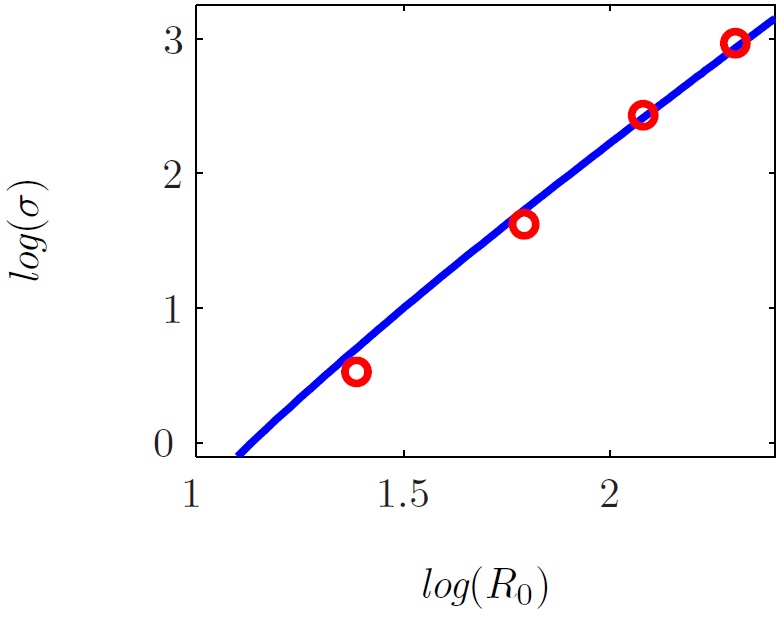}}}
	\caption{Envelope approximation functions (red dotted lines) and numerical solutions (blue solid lines) to $\rho(t)$ and $p_\bot(t)$ in the tPWM. The simulation parameters are $F=4$, $R_0=6$, $\gamma_0=10$, and $\eps=0.0004$. (\ref{fig:paper_MTC_M2_gamma=10_F=6_eps=0.0002}): Trapping cross-sections from analytical prediction (blue line) and numerical simulations (red circles) for $\gamma_0=10$, $\eps= 0.0002$, and $F = 6$ in the tPWM.}
	\label{fig:M2_460.20.000410}
\end{figure}

Two exemplary plots of the envelope functions $\tpyy(t)$ and $\tpyy(t)$ (red dotted lines) together with the simulated coordinate and momenta of a particle (blue solid lines) are shown in Fig.\ref{fig:M2_460.20.000410}. The analytical predictions are in good agreement with the simulations and converge very close to the actual enveloping functions.

To calculate the trapping conditions, we demand $|\rb|^2=(\xi^-)^2+\rho^2<R^2$ and show that for $n\gg 1$
\begin{equation}
	\xi_0^- < -\frac{2}{F} + \frac{R_0F^{3/4}\ct^{3/2}}{(F-R_0^2\eps)^{3/4}\g_0^{1/2}}\left(1.48 + 0.72\frac{R_0\eps\g_0^2 F}{2F-R_0^2\eps}\right)		\label{eqn:sufficient}
\end{equation}
is the condition that must hold to trap electrons. From Eq.(\ref{eqn:sufficient}) and $\rho_0^2=R(t)^2-(\xi_0^-)^2$ we calculate the trapping cross section $\sigma = \pi(R^2-\rho_{\min}^2)$ with $\rho_{\min}=\min \rho_0$. If we assume $\Delta R=R-\rho_{\min}=const$, we find that $\sigma(t)$ is linearly time-dependent. To confirm our estimation, we plot $\log(\sigma)$ and $\log(R_0)$ into one diagram (see Fig.\ref{fig:paper_MTC_M2_gamma=10_F=6_eps=0.0002}). As we observe, the analytical prediction (blue line) follows the numerically found trapping cross sections (red circles) and stays especially for large $R_0$ in a $5\%$ error range.\\

\begin{figure}[t]
	\subfloat[]{\label{fig:paper_AME_M2_448514_Energie_3Dquick_trap}\resizebox{5cm}{!}{\includegraphics{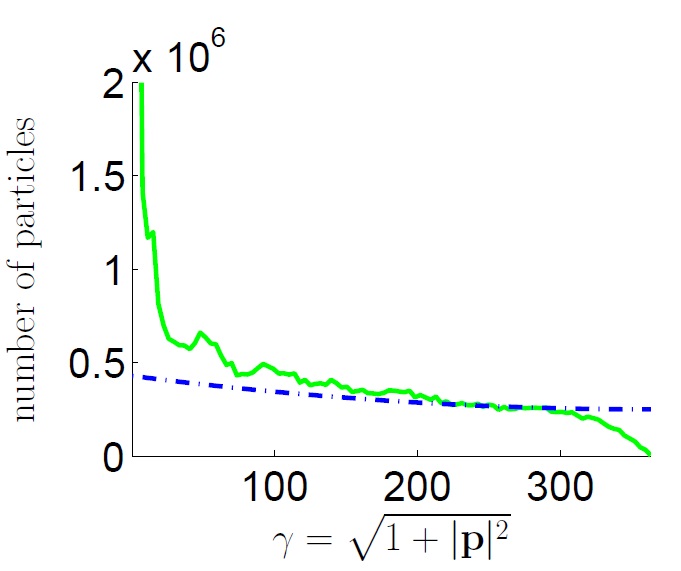}}}
	\hfill
	\subfloat[]{\label{fig:Gordienko2006}\resizebox{5cm}{!}{\includegraphics{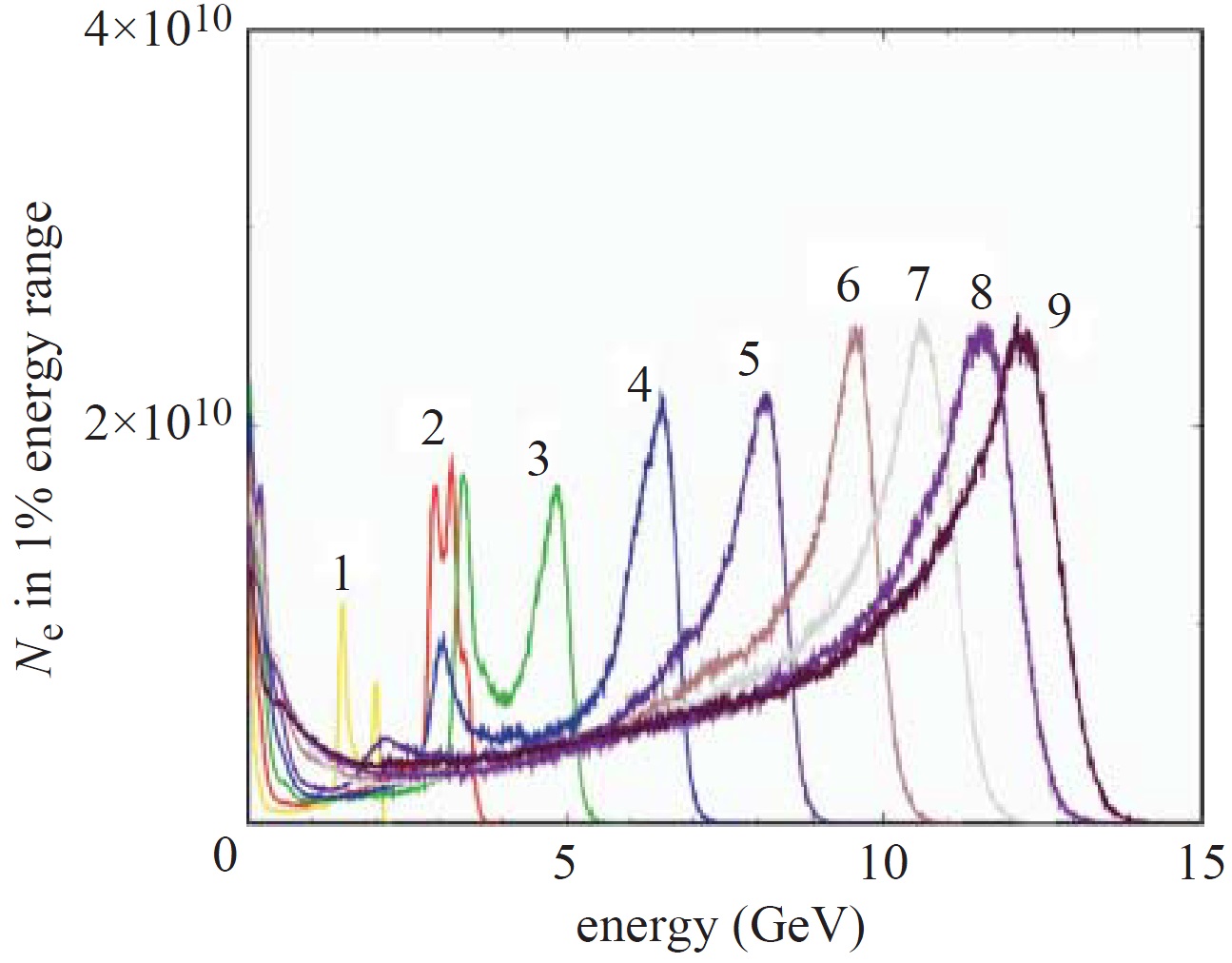}}}
	\hfill
	\subfloat[]{\label{fig:4815E}\resizebox{5cm}{!}{\includegraphics{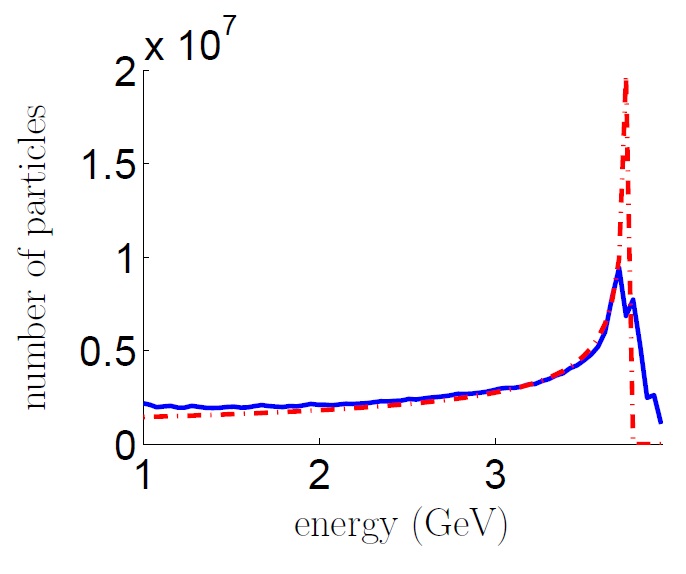}}}
	\caption{(\ref{fig:paper_AME_M2_448514_Energie_3Dquick_trap}): Energy distribution for $F=4$, $R_0=4$, $\eps=0.004$, and $\gamma_0=5$ in the tPWM. (\ref{fig:Gordienko2006}): PIC energy spectra taken from Fig.2 in \cite{Pukhov2006}. (\ref{fig:4815E}): Energy spectra in the hPWM for $\g_0=15$, $R_0=8$, and $F=4$.}
	\label{fig:ENERGY}
\end{figure}
To analyze the energy spectra in the tPWM a grid of electrons with zero momentum is placed in front of the bubble (see Fig.\ref{fig:Grid_2D}). Afterward, we solve the equations of motion for every individual particle, i.e. we neglect the interaction of the test electrons. The electron spectra we receive after the electron bunch of trapped electrons has passed its acceleration length look like that in Fig.\ref{fig:paper_AME_M2_448514_Energie_3Dquick_trap}. Here, we don't observe any mono energetic electron beams but a quasi equal distribution of the particles' energy. Apparently, this contradicts pic simulations results as shown in Fig.\ref{fig:Gordienko2006} and experiments \cite{Hidding2006,Mangles2004,Faure2004}. 

The monoenergetic spectra can be recovered in the full gradient model. Thus, in the following section we introduce another extension of the PWM that is still analytically treatable and preserves a production of mono energetic electron beams.\\

\textbf{The half-gradient PWM (hPWM)}\\
In the hPWM we still replace the perpendicular field $E_\bot$ by the constant field-strength parameter $F$. In longitudinal direction, however, we take the full field gradient. The resulting wake-field potential is
\begin{equation}
\Phi(\rb)=\frac{\xi^2+2F|\rho|}{4}-\frac{R_0^2}{4}.
\end{equation}
Here, the coordinate $\xi=z-V t$ always contains the group velocity of the generating laser pulse at the bubble front because the bubble radius is fixed. 

If we repeat the simulations of the electron grid in the the hPWM, we get energy spectra similar to those shown in Fig.\ref{fig:ENERGY}. As we see, both PIC-simulated spectra in Fig.\ref{fig:Gordienko2006} and our spectrum from a simulation with $F=4$, $R_0=8$, and $\gamma_0=15$ exhibit the same structure. Since the hPWM is able to describe the production of monoenergetic electron beams, we calculate the equations of motion
\begin{eqnarray}
\frac{dp_{||}}{dt} &= & -\Omega + \frac{p_\bot}{\g}\frac{\sgn(\rho)}F{4},													 	\label{eqn:SPWMEq_I}\\
\frac{dp_\bot}{dt} &= & -\left(1+\frac{p_{||}}{\g}\right)\frac{\sgn(\rho)F}{4},										 	\label{eqn:SPWMEq_II}\\
\frac{d\xi}{dt} 	 &= & \frac{p_{||}}{\g}-V, \hspace{0.5cm} \frac{d\rho}{dt}	= \frac{p_\bot}{\g},	\label{eqn:SPWMEq_III}
\end{eqnarray}
with $\Omega=(1+V)\xi/4$. A simple form for the envelope approximation functions to $\rho(t)$ and $p_\bot(t)$ can be found, if we express them in terms of $\tau=t-t_{max}$, where $\xi(t_{max})=0$. For the calculation we follow \cite{Kostyukov2010} again and assume $\g(t)\ll2(n_{cr}/n_0)^2$ for all $t$ for the electron energy. Then we have
\begin{eqnarray}
\tpxx(\tau) & \approx & p_{max} - \frac{(1+V)}{16\g_0^2}\tau^2				\\
\tpyy(\tau) & = & \sqrt{\Lambda}(F\tpxx(\tau))^{1/3},	\label{eqn:tpy}\\
\tyy(\tau)  & = & \Lambda(F\tpxx(\tau))^{-1/3},				\label{eqn:ty}
\end{eqnarray}
with $\Lambda = 2^{1/3}C(t_n)^2/8\Omega(t_n)^2$. As in the previous section, the constant $C(t_n)$ is used to fit our approximations to numerical simulations.

\begin{figure}[tbp]
	\centering
	\subfloat[]{\label{fig:51020PY}\resizebox{6cm}{!}{\includegraphics{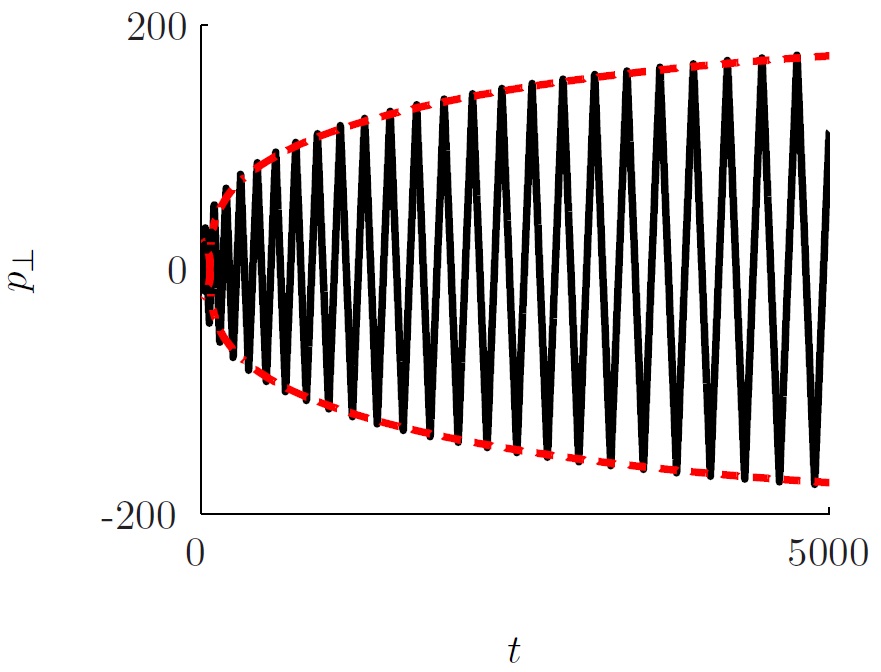}}}
	\hfill
	\subfloat[]{\label{fig:51020Y}\resizebox{6cm}{!}{\includegraphics{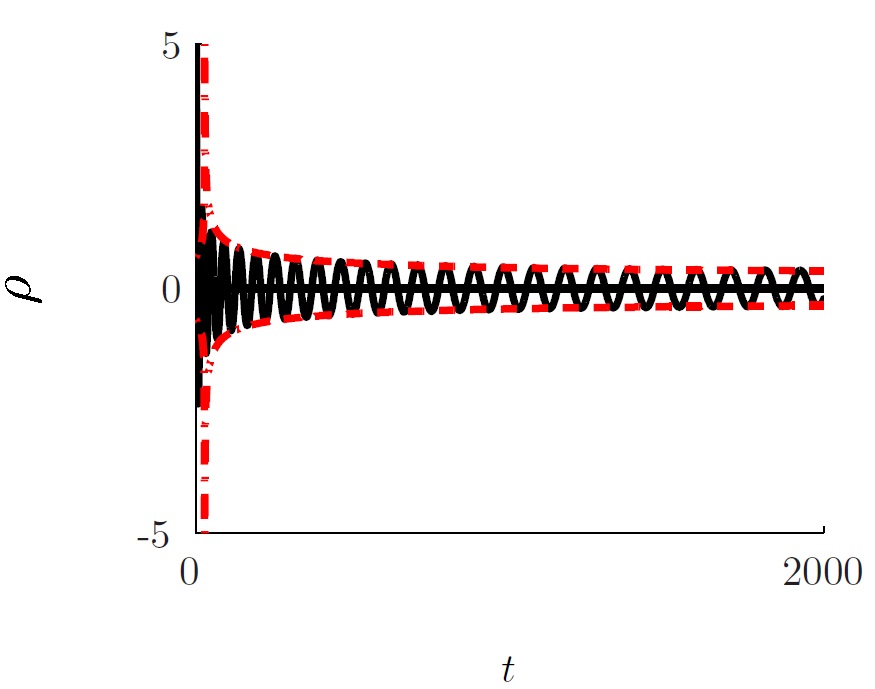}}}
	\caption{envelope approximation functions (red dotted lines) and numerical solutions (blue solid lines) to $\rho(t)$ and $p_\bot(t)$ in the hPWM. The simulation parameters are  $F=5$, $R_0=10$, and $\g_0=20$.}
	\label{fig:PYY}
\end{figure}

To compare the approximated functions to the actual amplitude envelope functions, we solve the equations of motion for a single electron with initial conditions $\xi_0^2+\rho_0^2 = R_0^2$, $ \pb_0=0$, and $\xi_0, \rho_0 >0$.  An exemplary plot of Eqs.(\ref{eqn:tpy}-\ref{eqn:ty}) together with numerical solutions is shown in Fig.\ref{fig:PYY}. The analytical predictions (red dotted lines) are in good agreement to the simulations (black solid lines) and converge very close to the actual enveloping functions.

\begin{figure}[t]
	\subfloat[]{\label{fig:sigma15}\resizebox{6cm}{!}{\includegraphics{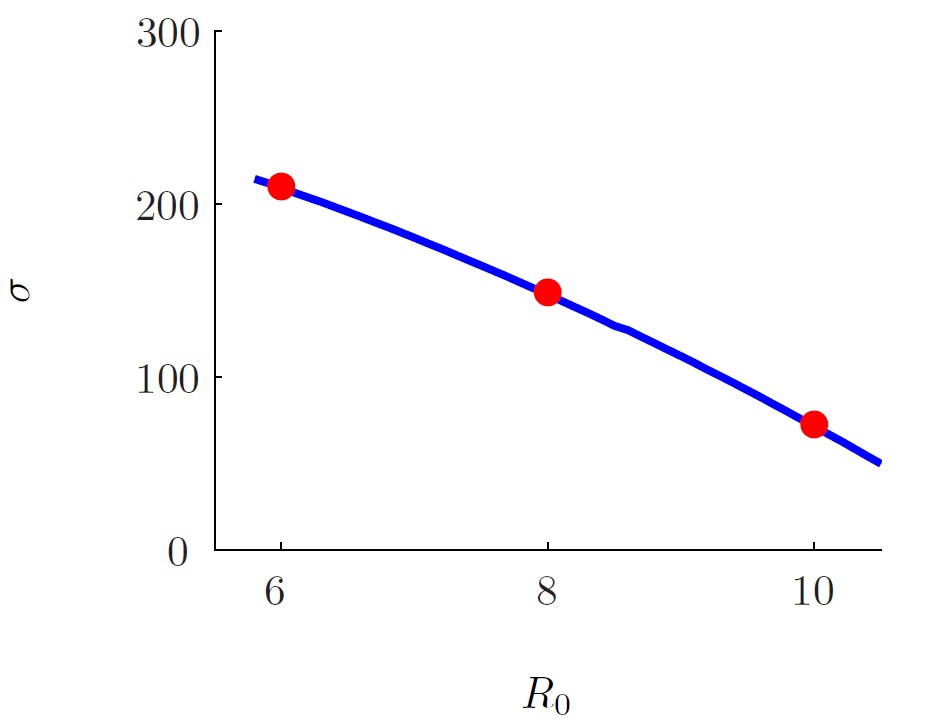}}}
	\hfill
	\subfloat[]{\label{fig:F15}\resizebox{6cm}{!}{\includegraphics{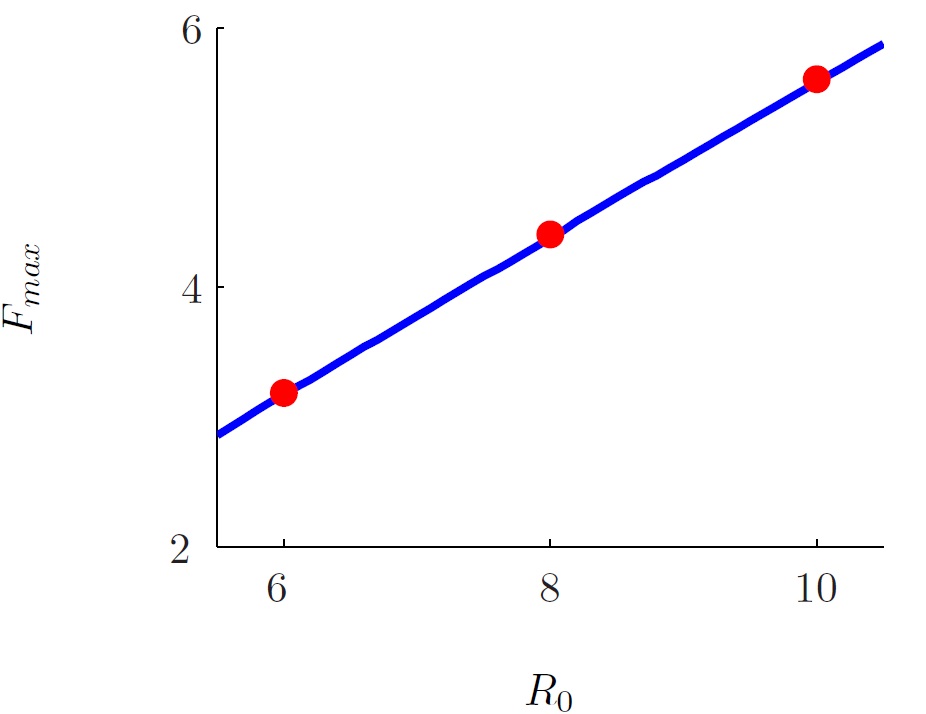}}}
	\caption{(\ref{fig:sigma15}): Trapping cross-section for $F=R_0/2$. (\ref{fig:F15}): Maximal applicable $F$. Both figures are simulated with $\gamma_0=15$ and belong to the hPWM.}
\end{figure}

Similar to Eqs.(\ref{eqn:sufficient}) the self-injection physics is described by the limiting condition
\begin{equation}
\rho_{\min} = F + \left(4 + F^2 - \frac{2P_0}{\g_0^2} - \frac{2\Lambda F^{2/3}}{P_0^{1/3}} + \frac{\Lambda^2}{(FP_0)^{2/3}}\right)^{1/2}	\label{eqn:rho}
\end{equation}
together with
\begin{equation}
P_0^{5/3} - \frac{\g_0^2\Lambda F^{2/3}}{3}P_0^{1/3} + \frac{\g_0^2\Lambda^2F^{-2/3}}{3} = 0.
\end{equation}
The trapping cross-section $\sigma = \pi(R_0^2-\rho_{\min}^2)$ is constant in time since $R_0$ and $\rho_{\min}$ are. If we plot $\sigma$ and $R_0$ into one diagram (see Fig.\ref{fig:sigma15}), we observe that for a large range of parameters the accuracy remains in a $1\%$ range.

To get some analytical expressions for the limits of the bubble parameters, the so called trapping conditions, we start again with Eq.(\ref{eqn:rho}) and demand $\rho_{\min}<R_0$. Then we find after some algebraic transformations
\begin{equation}
\g_0^2 < 2P_0(2FR_0 - R_0^2 + 4 - 2\Lambda F^{2/3}P_0^{-1/3} + \Lambda^2F^{-2/3}P_0^{-2/3})^{-1},
\end{equation}
as well as
\begin{equation}
F^{5/3} - \frac{\Lambda}{R_0P_0^{1/3}}F^{4/3} + \left(\frac{2}{R_0} - \frac{R_0}{2} - \frac{P_0}{R_0\g_0^2}\right)F^{2/3} - \frac{\Lambda^2}{P_0^{2/3}} < 0.
\end{equation}
These two equations give us all parameter limits for the hPWM. To demonstrate that these complicated-looking limits are indeed well calculated we plot the largest $F$ for which still an electron is trapped against $R_0$ in a simulation with one electron (compare former sections) in Fig.\ref{fig:F15}. As we see the analytical found $F_{max}$ (blue line) fits perfect to our simulations (red dots).

\textbf{Conclusion}\\
Summarizing our expansions of the original PWM we note two things. First, we see that it is possible to treat an enhanced interaction between the electron bunch and the bubble potentials. This in turn leads to a growing trapping cross-section and thus to a more efficient self-injection. 

Energy spectra in our simulations show that no mono energetic electron beams are generated, if all field gradients are neglected. To fix this problem, we invent a hybrid model that includes the right fields in moving direction and replaces the transverse field gradients by a field strength parameter. In this modification we observe the generation of mono energetic electron beams and are able to calculate trapping conditions analytically.

The next steps of our work will be to go back to the original full gradient model and to try to find the envelope approximation functions by other means than the introduction of a field strength parameter. If we succeed, trapping cross-sections and trapping conditions for the most physical model will be available without thinking about an arbitrarily set field strength parameter.\\

\textbf{Acknowledgment}\\
This work has been supported by the Deutsche Forschungsgemeinschaft via GRK 1203 and SFB TR 18.

\bibliography{Plasma_Physik}

\end{document}